# On the Interpretation of Spreadsheets within their Environment

Stephen Allen
ACBA (UK) LTD
steve@acba.co.uk

## ABSTRACT

*A demonstration in MS Excel to show how users can connect their spreadsheet models to the external environment that the model represents. We employ indexes to generate a list of relevant evidence that is hyperlinked to the context in which the evidence is discussed. The hyperlinks between the index and the contextual discussion have their own specific presentational identity. We contend that these presentational differences aid the integrity and understanding of complex models.*

*Where models are complex, separate individual results can lead to contradictory conclusions. The demonstration includes a methodology for interpreting the analyses within a workbook and presenting them in the form of a standard written report.*

## 1 INTRODUCTION

In our Microsoft, business user centric world, we have to make a big decision early on. Are we going to treat Excel as simply one element of the total business package and swap between it and the other packages (Word and PowerPoint mainly)? Alternatively we could treat Excel as a standalone package which holds all the primary logic and background rationale for our process, its methodological approach and outputs. In this presentation we have opted for the standalone version.

Most spreadsheets are designed to undertake a specific job within a specific (and often fairly confined) environment. In the standalone context we must create connections with the outside world. Failure to make these connections opens the risk of misinterpreting the nature and relevance of the workbook results. It is these connections and the methodologies available for presenting these connections that form the basis of this presentation.

We consider some practicable processes for presenting those connections and a methodological approach for creating the necessary processes. These processes all employ the hyperlink as the basic mechanism. The methodological differences in the use of hyperlinks considered here are aimed entirely at presentation. Our contention is that these presentational differences aid the integrity and understanding of complex models in spreadsheets.

This presentation considers creating connections between the spreadsheet and the external environment in two primary contexts:

- Documentary evidence held in electronic form which forms the underpinning basis for the spreadsheet analysis
- Evidence of the arguments generated by the spreadsheet user/creator in support of a particular conclusion

## 2 IMPORTANCE OF LISTS AND INDEXES

The creators of Wikipedia and anyone preparing a scientific paper know the importance of generating a list of sources. Users of Word and WordPerfect will be well aware of the sophisticated methods in those programs for creating such reference/source lists. There is no equivalent in Excel.

Nor can we identify any published research on the matter, but two recent papers have discussed issues associated with 'quality' as applied to spreadsheets. Balson [2010] refers to '*laying models out clearly, documenting them fully, [and] keeping them simple*'. The emphasis here is on clear source documentation. In their seminal paper "Towards Evaluating the Quality of a Spreadsheet" Grossman et al [2011] propose a principle that '*Spreadsheet quality should be evaluated in terms of the spreadsheet artefact, not the process.*' In the context of this paper, the artefact includes the sources that validate it. It answers the question "How do we know that this model (and its input) is correct?"

The need for such source lists in Excel remains open to argument, but the ACBA software [ACBA, 2002] for creating these relatively complex workbooks assumes that there is an underpinning need for establishing the relationship between the workbook analysis and the real world. The type of list we create has superficial similarities to a bibliography of references in an academic paper, but is much less rigid in the scope of its usage.

The primary similarity concerns the expectation that the reference will appear in context, as well as within a list of sources. So under our main software convention [ACBA, 2002] a user can only create an indexed source reference from within a worksheet created under the control of the company software. This is what establishes the link between context and the reference. The reference is not only visual. There are active bi-directional hyperlinks between the list of references and the context in which it is employed.

The counterargument is that the process of creating a separate list generates a barrier between the reader and the source material. Why not generate a direct hyperlink between the data in the spreadsheet and its source? This is perfectly practicable and would lessen the barrier between reader and source. However, the context of this proposal is that, frequently, there will be multiple sources. The completeness of the sources employed, the relationships between them and, consequently, the model's overall quality, will often be a significant issue. A reference listing that brought all the source material together was deemed a priority.

A further counterargument concerns the use of separate hyperlink references. The proposal here suggests that another barrier is created by linking to a reference (like **F051/1**) rather than directly to the text itself. We accept that this is a notional barrier however the stylized reference allows us to generate a logical order to the list of references / bibliography such that related items of evidence are posted together.

## 3 SOURCE MATERIAL

The pace of change from paper storage of data to electronic storage of data over the last 30 years has been astonishingly rapid. This change has created an entirely new environment associated with the security and integrity of that data. The ACBA software sought to offer a methodology for both indexing and securing the integrity of associated source material.

Under the process, a user creates a reference to the source material from within the appropriate worksheet (or working paper). This generates an indexed entry on the Index of Evidence (or bibliography). The user is then invited to identify the source of the evidence (assuming that it is an electronic file). This is copied into a directory structure below the primary working file for security purposes and a hyperlink generated to this copied version of the original source. The hyperlink is deliberately generated in such a way that if the working paper control file is moved from its original location, the hyperlinks in the Index of Evidence fail.

The Index of Evidence (Figure 1) offers several layers of information. It provides:

- a brief description of the evidence itself
- a hyperlink to a copy of the original source
- a hyperlink to the working paper that refers to or analyses that source material
- a contextual appreciation of the importance of the evidential material to the whole analysis

**Figure 1 – The Documentary Evidence Index [ACBA-EWP Help, 2009]**

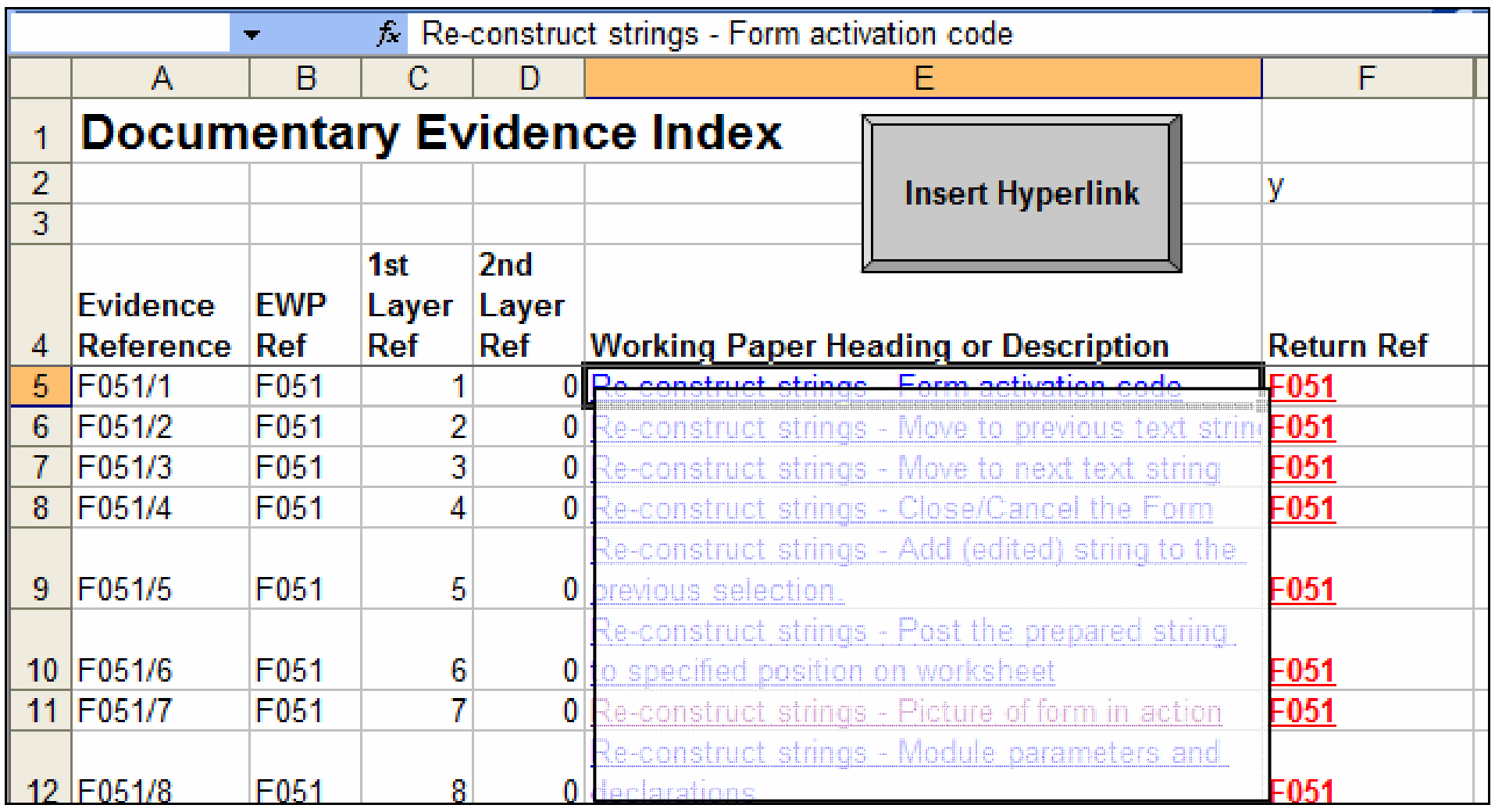

Re-construct strings - Form activation code

| | A | B | C | D | E | F |
|---|---|---|---|---|---|---|
| 1 | Documentary Evidence Index | | | | Insert Hyperlink | |
| 2 | | | | | | y |
| 3 | | | | | | |
| 4 | Evidence Reference | EWP Ref | 1st Layer Ref | 2nd Layer Ref | Working Paper Heading or Description | Return Ref |
| 5 | F051/1 | F051 | 1 | 0 | Re-construct strings - Form activation code | F051 |
| 6 | F051/2 | F051 | 2 | 0 | Re-construct strings - Move to previous text strin | F051 |
| 7 | F051/3 | F051 | 3 | 0 | Re-construct strings - Move to next text string | F051 |
| 8 | F051/4 | F051 | 4 | 0 | Re-construct strings - Close/Cancel the Form | F051 |
| 9 | F051/5 | F051 | 5 | 0 | Re-construct strings - Add (edited) string to the previous selection. | F051 |
| 10 | F051/6 | F051 | 6 | 0 | Re-construct strings - Post the prepared string to specified position on worksheet | F051 |
| 11 | F051/7 | F051 | 7 | 0 | Re-construct strings - Picture of form in action | F051 |
| 12 | F051/8 | F051 | 8 | 0 | Re-construct strings - Module parameters and declarations | F051 |

The demonstration will show (very briefly) how to create the working paper reference and the methodology for posting the viewable hyperlink within the Documentary Evidence Index.

## 4 EVIDENCE BASED CONCLUSIONS

Dinmore [2009] states that "*in traditional programming, a number of mechanisms have been employed to capture programmer knowledge, principally various forms of documentation internal and external to the source code*". Louise Pryor [2006] offers a slightly broader perspective described as *a brief characterisation of the purposes and forms of documentation in and of spreadsheets*. Both these papers provide a sound approach to the documentation of the design, structure and purpose of a spreadsheet, but they do not expand on the interpretation of results.

Not all spreadsheet analyses generate simple numeric or logical conclusions. The spreadsheet paradigm itself, to some extent, encourages the prospect of interpreting the factual analyses in context (c.f. Excel's cell comment function). Some researchers have sought to expand the use of the cell comment function to give a broader view of the context of individual cells [Payette, 2006]. This appreciation of context has been discussed in previous papers [Pryor 2003, Banks and Monday 2002] where there is a recognition that the potential for the misinterpretation of a spreadsheet is a serious issue.

Since an individual's background experience (or even social/religious belief) can vary hugely, the conclusions we draw from a common set of facts can be remarkably different. This variability of interpretation applies to business analysis, audit and a range of other analytical approaches beyond the strict confines of the accounting world. We take the view that it entirely excludes scientific analyses from whatever discipline, but even this may be open to argument.

For example, given a complex and wide-ranging set of analyses in support of an internal audit review, the reviewer may generate a series of conclusions/opinions. Some of these may well conflict with each other. The ACBA software provides a fairly sophisticated process through which the project owner can validate and justify his or her opinions. The software also allows the project owner to establish the relationships and relative importance between conflicting findings.

### *4.1 'Controlled Statements'*

The process is based on a company developed 'object', more strictly a family of 'objects', known as 'Controlled Statements' [ACBA, 2003]. These 'Controlled Statements' are divided into types, each with its own specific characteristics/properties. The characteristics determine the potential relationship that each 'Controlled Statement' has to another within the family – see Figure 2. We use these properties to establish a logical order for the presentation of an argument. The role of the owner is to connect the 'Controlled Statements' together to form a chain that represents the logic of the case statement.

**Figure 2 - A linked chain of 'Controlled Statements'**

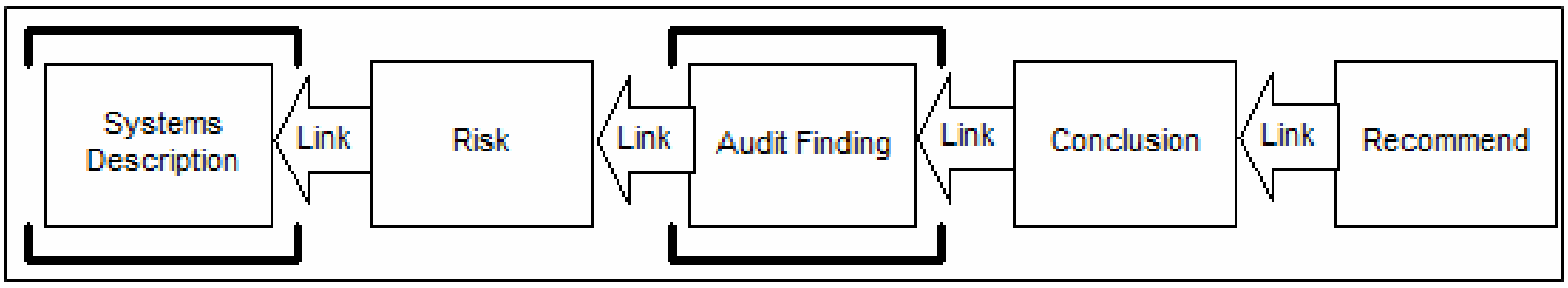

'Controlled Statements' are deliberately designed to look separate and different from the primary analysis or model of the workbook (Figure 3). Like Excel's cell comments they have a standardised format. But this object is significantly more sophisticated than a cell comment. 'Controlled Statements' can form a linked chain of commentary. The forward links are coloured lime green and the rearward facing links purple.

**Figure 3 - A 'Controlled Statement' looks separate and different**

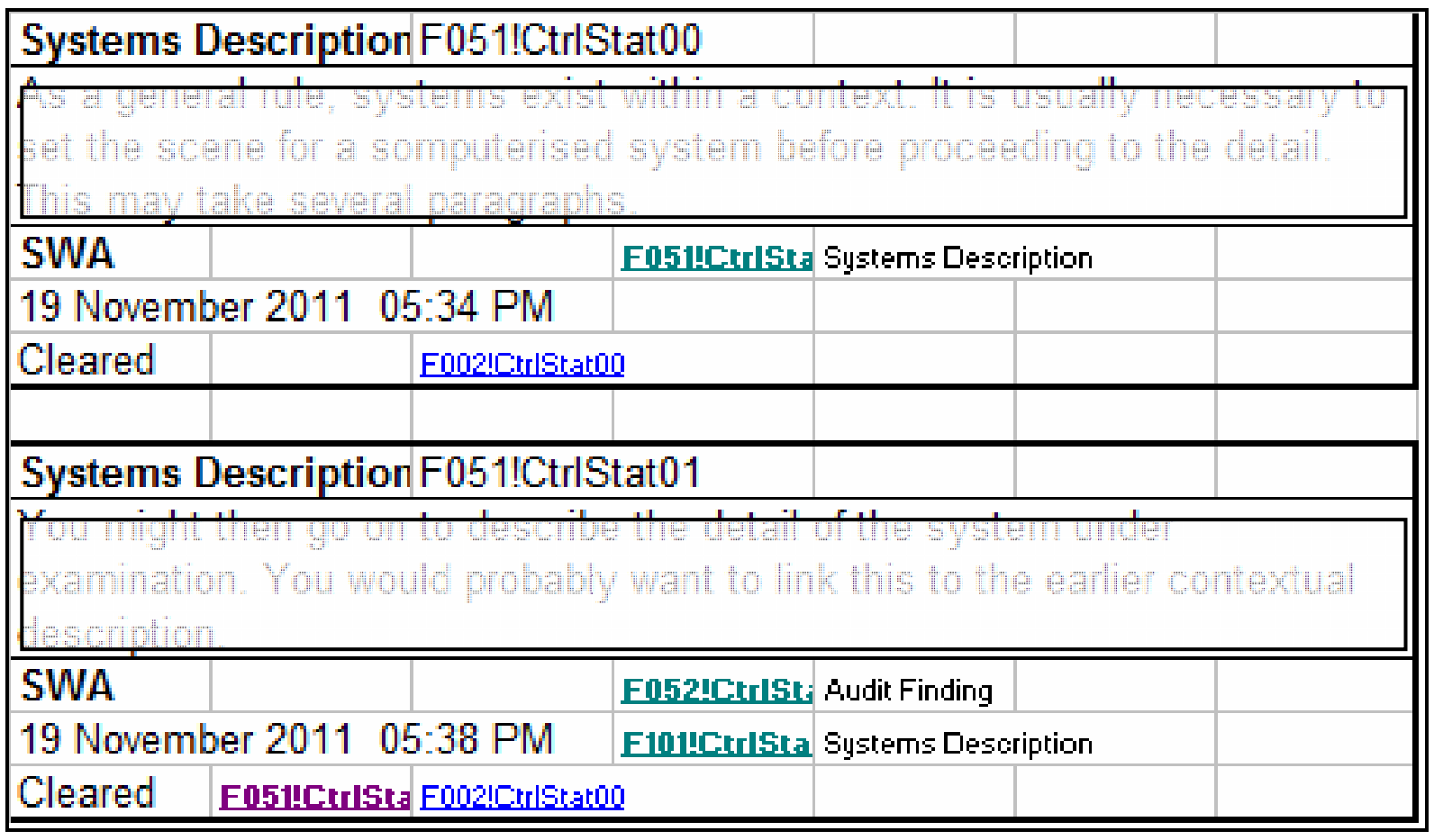

One of the unusual characteristics of 'Controlled Statements' is that their management methodology allows them to be located adjacent to (or at least fairly close to) the source data or analytical data that they represent. This is intended to provide instant and incontrovertible support for the statement. It is the linguistic equivalent of a logical assertion.

When considering a workbook containing multiple sheets, linked statements can be very distant from each other. The expectation of the linkage series is that it creates a single strand of logic related to a specific argument.

### *4.2 Module Layer Review*

A project may contain many such strands of logic in support of an overall conclusion or opinion. The role of the software is to provide a methodology for the owner to bring the separate arguments together, so that he or she can present a comprehensive case for a specific view or conclusion.

A special process pulls each of the linked 'Controlled Statement' together into a logical order within a simple readable view – see Figure 4. In the ACBA parlance, this is known as 'module layer review'. It provides a crucial opportunity for the user to review his own commentary away from the source data, and take account of the loss of visual impact that this will have on a reader who does not know the details of the system or the results on which the commentary is based.

This review process allows the commentator to review his own commentary

- For sense and clarity
- To ensure that each major concept is separately and uniquely headed
- To examine the relative importance of each major concept to the overall presentation of a coherent logic for the system in question
- To omit those concepts that no longer appear relevant to the overall presentation.

**Figure 4 - Drawing individual 'Controlled Statements' together**

| | | |
|---|---|---|
| 5 | | |
| 6 | **Describing the Primary System** | |
| 7 | Systems Description | F051!CtrlStat00 |
| 8 | As a general rule, systems exist within a context. It is usually | |
| 9 | necessary to set the scene for a somputerised system before | |
| 10 | proceeding to the detail. This may take several paragraphs. | |
| 11 | | |
| 12 | Systems Description | F051!CtrlStat01 |
| 13 | You might then go on to describe the detail of the system under | |
| 14 | examination. You would probably want to link this to the earlier | |
| 15 | contextual description. | |
| 16 | | |
| 17 | Audit Finding | F052!CtrlStat0 |
| 18 | Usually this involves, interpreting some form of spreadsheet | |
| 19 | anaysis. Within the audit emvironment we would refer to this | |
| 20 | as an audit finding | |
| 21 | | |
| 22 | Audit Finding | F053!CtrlStat0 |
| 23 | We rarely get a usable result in the first phase. We might try | |
| 24 | a more refined approach. | |
| 25 | | |
| 26 | Conclusion | F001!CtrlStat00 |
| 27 | Our conclusion from the testing of the primary is X | |
| 28 | | |
| 29 | Systems Description | F101!CtrlStat00 |
| 30 | Let us assume that we have a secondary system that is very closely | |
| 31 | related to our primary one. | |
| 32 | | |
| 33 | Audit Finding | F102!CtrlStat00 |
| 34 | We test this secondary systems to verify both its relationship | |
| 35 | to the primary systems and the validity of the results delivered. | |
| 36 | | |
| 37 | Conclusion | F001!CtrlStat01 |
| 38 | Our conclusion from the testing of the secondary systems is Y. | |
| 39 | | |
| 40 | **SWA** | |
| 41 | 20 November 2011 06:40 AM | |
| 42 | Report | |
| 43 | | |

## 4.3 *Create Report Control Sheet*

The final phase for the commentator is to create a report control sheet. This allows the user to control the order of presentation of the elements of the overall argument and set the arguments in context employing up to three levels of heading.

**Figure 5 – Control Sheet for Editing the Overall Structure of a Report**

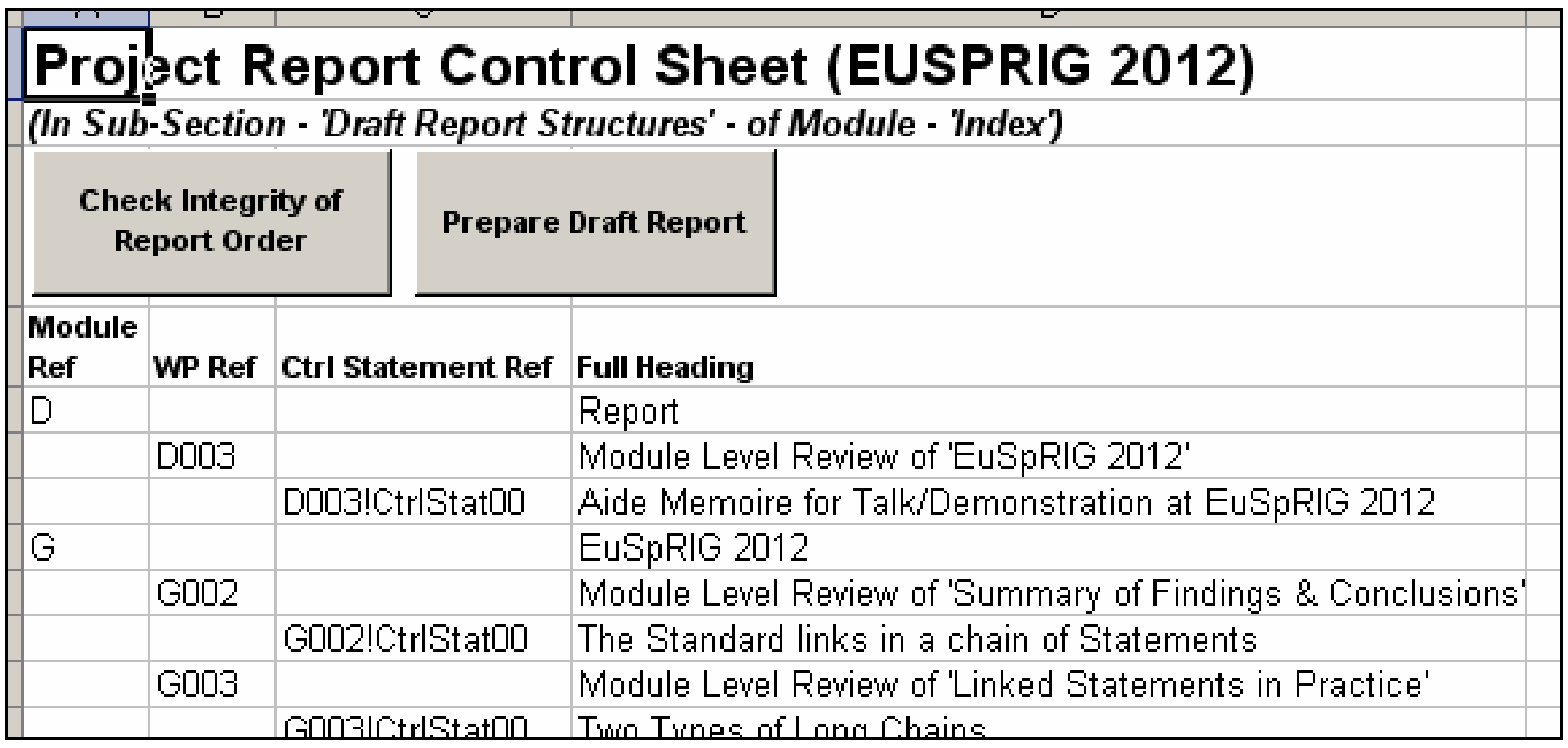

**Project Report Control Sheet (EUSPRIG 2012)**

***(In Sub-Section - 'Draft Report Structures' - of Module - 'Index')***

Check Integrity of Report Order | Prepare Draft Report

| Module Ref | WP Ref | Ctrl Statement Ref | Full Heading |
|---|---|---|---|
| D | | | Report |
| | D003 | | Module Level Review of 'EuSpRIG 2012' |
| | | D003!CtrlStat00 | Aide Memoire for Talk/Demonstration at EuSpRIG 2012 |
| G | | | EuSpRIG 2012 |
| | G002 | | Module Level Review of 'Summary of Findings & Conclusions' |
| | | G002!CtrlStat00 | The Standard links in a chain of Statements |
| | G003 | | Module Level Review of 'Linked Statements in Practice' |
| | | G003!CtrlStat00 | Two Types of Long Chains |

In Figure 5 – the order of the presentation of arguments (those that are identified under 'Ctrl Statement Ref') can be moved and the order of presentation altered to suit the commentator's view of the logic.

Similarly values under the columns entitled 'Module Ref' and 'WP Ref' represent first and second level headings respectively. The heading texts can be edited and their position moved (or even deleted entirely) to suit the commentator's view of the desired structure of the output report.

## 5 DEMONSTRATION

The process takes place in several phases which are:

- Create and link 'Controlled Statements' distributed throughout the project.
- Bring them together and present in a logical (readable) order
- Create an order and structure for the overall case you wish to present
- Print out in MS Word

The demonstration at EuSpRIG 2012, which presented the practical application of this paper, was delivered through an MS Excel workbook generated under the control of the ACBA Electronic Working Papers software [ACBA, 2002]. The project file itself illustrates the active nature of the software control mechanisms. A single project may well be active over an extended time period. In contrast, the software that generates 'Controlled Statements', is intended to deliver a snapshot of concepts and opinions at a specific point in time.

This demonstration project [Allen, 2012] records the approach taken at two conferences which considered the role of 'Controlled Statements': EUSPRIG 2012 and one delivered in January 2012 [XLDevCon]. The demonstration project file is divided into two indexed levels – Modules and Sub-Sections. The elements of the file that concern this demonstration specifically are as follows

**Table 1 – Relevant Parts of the Demonstration Control workbook [Allen, 2012]**

| Ref | Module | Ref | Sub-Section | Associated Working Papers |
|---|---|---|---|---|
| A | Index | A050 | Draft Report Structures | Project Report Control Sheet (EUSPRIG 2012) – A052 |
| D | Report | D000 | EUSPRIG 2012 | Aide Memoire for Talk/Demonstration at EuSpRIG 2012 – D001 |
| | | | | The Case for a New Class of 'Controlled Statement' - Method or Detailed Audit Plan – D002 |
| | | | | Module Level Review of 'EuSpRIG 2012' – D003 |
| G | EUSPRIG 2012 | G000 | Summary of Findings & Conclusions | The Standard links in a chain of Statements – G001 |
| | | | | Module Level Review of 'Summary of Findings & Conclusions' – G002 |
| | | | | Module Level Review of 'Linked Statements in Practice' – G003 |
| | | G100 | Linked Statements in Practice | Two Types of Long Chains – G101 |
| | | | | Some 'Controlled Statements' must have a Parent – G102 |
| | | | | Branched Chains – G103 |

VBA macros applied to the 'Project Report Control Sheet (EUSPRIG 2012)' generated a draft report in MS Word [ACBA, 2012] – see Figure 6.

The draft report depicted in Figure 6 is not expected to be publishable as is. The output direct from the operational Excel workbook is necessarily condensed and may contain terse observations related to specific Excel artefacts (e.g. ranges, tables, graphs etc). It will need some reshaping for client presentation even though much improved clarity can be achieved through the module layer review phase (paragraph 4.2).

**Figure 6 – Example of Draft Report Output in MS Word**

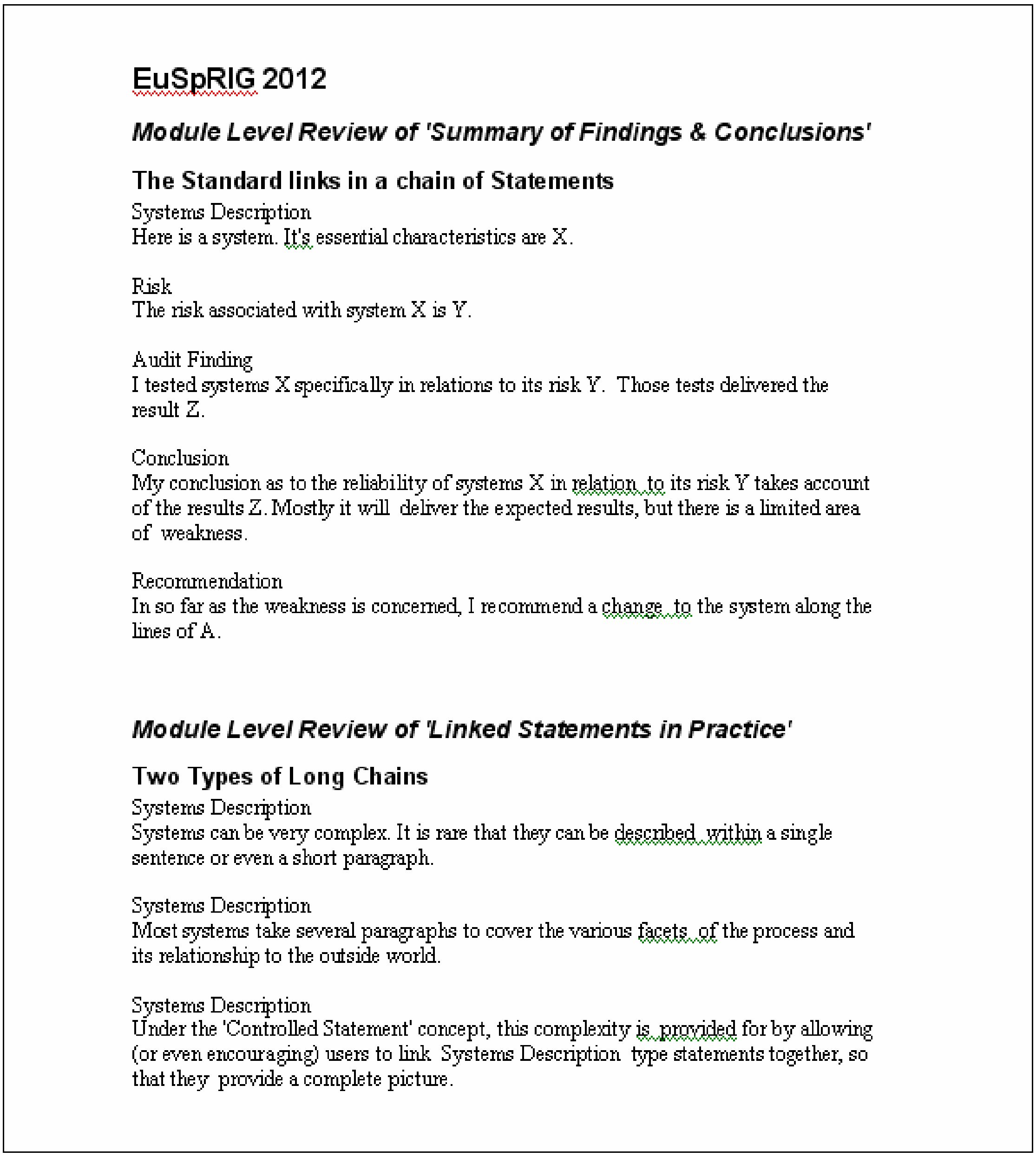

EuSpRIG 2012

***Module Level Review of 'Summary of Findings & Conclusions'***

**The Standard links in a chain of Statements**

Systems Description
Here is a system. It's essential characteristics are X.

Risk
The risk associated with system X is Y.

Audit Finding
I tested systems X specifically in relations to its risk Y. Those tests delivered the result Z.

Conclusion
My conclusion as to the reliability of systems X in relation to its risk Y takes account of the results Z. Mostly it will deliver the expected results, but there is a limited area of weakness.

Recommendation
In so far as the weakness is concerned, I recommend a change to the system along the lines of A.

***Module Level Review of 'Linked Statements in Practice'***

**Two Types of Long Chains**

Systems Description
Systems can be very complex. It is rare that they can be described within a single sentence or even a short paragraph.

Systems Description
Most systems take several paragraphs to cover the various facets of the process and its relationship to the outside world.

Systems Description
Under the 'Controlled Statement' concept, this complexity is provided for by allowing (or even encouraging) users to link Systems Description type statements together, so that they provide a complete picture.

## 6 CONCLUSION

This demonstration concentrates on the control and presentation of evidence within a relatively secure environment. There is not always a need for such security. The hyperlinks in the Evidence Index could equally link to publicly available sources over the internet. The strength of the indexed approach lies in the perspective that it offers a reviewer regarding the scope and detail of the project.

The requirement to argue the merits of conflicting or multiple strings of an argument contained in an Excel workbook is not an every day occurrence. The overall advantage of this methodology lies in the degree of automation that generates its traceability back to the underpinning evidence on which the argument is based.

## REFERENCES

ACBA (2002) “Electronic Working Papers” http://www.acba.co.uk/WP.htm 08.00pm 24/07/2012

ACBA (2003) “Controlled Statements” http://allen.netcom.co.uk/DownLoads/ACBA%20Controlled%20Statements.doc 08.00pm 24/07/2012

ACBA (2012) “An automatic report generated in MS Word from ‘Controlled Statements’ within an ACBA-EWP (MS Excel) project file.” http://acba.co.uk/DownLoads/CRCS_Output.doc 08.00pm 24/07/2012

ACBA-EWP Help (2009) “Documentary Evidence Index” http://allen.netcom.co.uk/EWP-Help/_2370src9d.htm 08.00pm 24/07/2012

Allen, S. W. (2012) “Aide Memoire for Talk/Demonstration at EuSpRIG 2012” http://acba.co.uk/DownLoads/ExcelDevSeminat.xls#D001!A1 08.00pm 24/07/2012

Balson, Dermot (2010) “Changing User Attitudes to Reduce Spreadsheet Risk” Proc. European Spreadsheet Risks Int. Grp. (EuSpRIG) 2010 133-137” arXiv:1009.5701v1

Banks, D., Monday, A. (2002) “Interpretation as a Factor in Understanding Flawed Spreadsheets” Proc. European Spreadsheet Risks Int. Grp. 2002 13 21 ISBN 1 86166 182 7 arXiv:0801.1856v1

Dinmore, M. (2009) “Documenting Problem-Solving Knowledge: Proposed Annotation Design Guidelines and their Application to Spreadsheet Tools” Proc. European Spreadsheet Risks Int. Grp. (EuSpRIG) 2009 57-68 ISBN 978-1-905617-89-0 arXiv:0908.1192v1

Grossman, T. A., Mehrotra, V., Sander, J. C. (2011) “Towards Evaluating the Quality of a Spreadsheet: The Case of the Analytical Spreadsheet Model” Proc. European Spreadsheet Risks Int. Grp. (EuSpRIG) 2011 ISBN 978-0-9566256-9-4 arXiv:1111.6907v1

Payette, R. (2006) “Documenting Spreadsheets” Proc. European Spreadsheet Risks Int. Grp. (EuSpRIG) 2006 163-173 ISBN:1-905617-08-9 arXiv:0803.0165v1

Pryor, L (2003) “Correctness is Not Enough” Proc. European Spreadsheet Risks Int. Grp. (EuSpRIG) 2003 117-122 ISBN 1 86166 199 1 arXiv:0808.2045v1

Pryor, L (2006) “What’s the point of documentation?” Proc. European Spreadsheet Risks Int. Grp. (EuSpRIG) 2006 arXiv:1011.1021v1

XLDevCon (2012) “Controlled Statements within ACBA Electronic Working Papers” http://allen.netcom.co.uk/DownLoads/ExcelDevSeminar-Jan2012-Revised.ppt 08.00pm 24/07/2012